\documentclass[epjCONF]{svjour}
\usepackage{amssymb}
\usepackage{amsmath}
\usepackage{graphics}
\usepackage[varg]{txfonts} 
\usepackage[latin1]{inputenc}

\newcommand{\nn}{\nonumber\\}
\renewcommand{\d}{\partial}

\newcommand{\rh}{\varrho}
\newcommand{\exv}[1]{\left\langle{#1}\right\rangle}
\newcommand{\ep}{\varepsilon}

\renewcommand{\k}{{\bf k}}
\newcommand{\p}{{\bf p}}
\newcommand{\Tr}{\mathop{\textrm{Tr}}}
\newcommand{\pint}[2]{{\int\!\frac{d^{#1}#2}{(2\pi)^{#1}}\,}}
\newcommand{\bra}[1]{\left\langle{#1}\right|}
\newcommand{\ket}[1]{\left|{#1}\right\rangle}
\session-title{Hot and Cold Baryonic Matter -- HCBM 2010}
\begin{document}
\title{Transport coefficients in non-quasiparticle systems}
\author{Antal Jakov\'ac\thanks{\email{jakovac@phy.bme.hu}}}
\institute{Insitute of Physics, BME Technical University, H-1111
  Budapest, Hungary}
\abstract{ Transport coefficeints, in particular the shear viscosity
  to entropy density ratio is studied in systems where the small-width
  quasiparticle assumption is not valid. It is found that $\eta/s$ has
  no unversal lower bound, the minimal value depends on the system and
  the temperature, and can be even zero. We construct models where the
  $1/4\pi$ conjectured bound is violated.  } 
\maketitle
\section{Introduction}
\label{intro}

Recent results from RHIC \cite{RHIC} suggest that hadronic matter near
$T_c$ behaves as an almost ideal fluid \cite{Shuryak}. 
Matter with small viscosity is very hard to described theoretically,
based on model calculations. The main reason is that the usual
perturbative approach expands the theory around the free gas limit,
where, by construction, the elementary excitations are in the ballistic
regime with infinite mean free path. Since all transport coefficients
are proportional to the mean free path, we obtain infinite results for
them in the unperturbed regime. With formulae we can state that
$\eta\sim 1/\sigma \sim 1/g^4\times \mathrm{logs}$ in QCD, where
$\sigma$ is the cross section, $g$ is the coupling constant. In fact
transport description assumes diffusion, the transport coefficients
are diffusion constants. If we enforce a diffusive description to the
ballistic regime, we obtain infinite diffusion constant.

Therefore the small shear viscosity coefficient measured at RHIC means
that the elementary excitations of the matter after the collision have
very short mean free path. Short mean free path means short lifetime,
large width, and so a not a small-width quasiparticle system. If we
still want to enforce an interacting free gas picture, it can be done
only with very strong interactions. To describe such a system requires
new approches.

A possible exact method would be using lattice Monte Carlo (MC)
simulations. One can measure $\exv{ T_{12}(x) T_{12}(0) }$ correlator
on lattice in Euclidean time, and use exact integral relations
\cite{MCeta} to determine the commutator
$C(x)=\langle[T_{12}(x),T_{12}(0)]\rangle$. One has to invert this
relation based on only a discrete set of time arguments. To make it
feasible one has to use prior knowledge on the result
(eg. $C(\omega>0)>0$), and incorporate it via the Maximal Entropy or
similar method. Then one can extract the shear viscosity from the Kubo
formula as $\eta = \lim\limits_{\omega\to
  0}\frac{C(\omega,\k=0)}{\omega}$. The hard point of this analysis is
that the Euclidean correlation function depends only very weakly on
the derivative of $C$ at zero frequency, ie. viscosity
\cite{Petr}. This means that one has to have large systematical
uncertainties. Nevertheless one can give estimates as
$\eta/s=0.102\,(56)$ at $T=1.24T_c$ \cite{MCeta}, where the
uncertainty reflects the statistical error.

Another method to treat strongly interacting models to find a dual
description with small coupling. For QCD such dual theory is not
known, the ``closest'' theory where a dual theory can be constructed
is the ${\cal N}=4$ super Yang-Mills (SYM) theory. If the coupling and
the number of colors are infinite ($N_c\gg1$ and
$\lambda=g^2N_c\gg1$), then the dual model is a five dimensional
gravity with AdS spacetime. Then, from graviton scattering, one can
obtain the shear viscosity coefficient and $\eta/s$, the latter turns
out to be $1/4\pi$ \cite{KSS}. For a wide class of theories this is a
lower bound \cite{Bucheletal}.

Although strictly speaking there are no exact perturbative methods to
treat the strongly interacting matter, one can give arguments to
estimate $\eta/s$ \cite{DG}. This argumentation uses the quasiparticle
picture: there $\eta\sim\epsilon \tau$ and $s\sim n$ where $\epsilon$
is the energy density, $\tau$ is the quasiparticle lifetime and $n$ is
the particle density. In this case $\eta/s\sim E\tau$ where $E$ is the
particle energy. For quasiparticles $E\ge\Delta E$, and so $\eta/s \ge
\Delta E\tau \gtrsim \hbar$. Since the so obtained value is close to
the one given by the ${\cal N}=4$ SYM at large couplings, it is
tempting to interpret $1/4\pi$ as a universal lower bound for any
existing matter.

This strong statement has been discussed a lot in the recent
literature, and there are counterarguments given against its generic
validity. In the 5D gravity models, one can construct higher curvature
and dilaton models \cite{AdS1} where the $1/4\pi$ is no more a lower
bound. An explicit couterexample was constructed in \cite{Cohan},
where the authors use a multicomponent model where the mixing entropy
can tune the $\eta/s$ to zero.

As the argumentation of \cite{DG} is concerned, it is strongly based
on quasiparticle picture. In an interacting model the spectral
function always contains other structures than the quasiparticle peak,
and these can change the conclusions based on quasiparticles. For
example the presence of a multiparticle continuum decreases
considerably the $\eta/s$ ratio \cite{NNG}.

From the experimental side, one can analize the anisotropy of the
created particles, in particular the $v_2$ coefficient. From hydro
fits one can readily give an upper bound for the $\eta/s$ ratio
\cite{RR} ($\eta/s<0.16$), but it is much more difficult to present a
lower bound. Taking into account the quadratic $p_T$ dependence of
$\eta/s$ \cite{Teaney}, the true value can be very close to the
conjectured lower bound \cite{RL}.

In other, more controllable experiments it is found \cite{VK} that
supercritical fluids may have very small $\eta/s$ (or physically
equivalent) ratio.

All these spectulations reflect the fact that the status of the
minimum of the $\eta/s$ ratio is still not clarified. The goal of the
present work is to determine, what can be said about this ratio in a
generic model, without assuming a small-width quasiparticle
picture. The generic model will be defined by giving the energy levels
of the system (density of states, DoS). For a more detailed discussion
see Ref.s \cite{jak0} and \cite{jak}.

The paper is organized as follows. In Section \ref{sec:1} we analyze
the $\eta/s$ ratio in a generic model and determine the minimum of
it. In Section \ref{sec:2} we consider some models where the
conjectured $1/4\pi$ bound can be violated. We close the discussion
with conclusions in Section \ref{sec:concl}.

\section{Transport and entropy in generic models}
\label{sec:1}

A definite difference between the quasiparticle DoS and a generic one
is the presence of continuum, ie. there is no dispersion relation, at
fixed spatial momentum the system has still a lot of energy levels. To
give an account for this property we define
\begin{equation}
  \sum_n \ket n\bra n = V\sum_{\cal Q}\!\! \pint 4p\, \rh_{\cal Q}(p)
  \ket{p,{\cal Q}}\bra{p,{\cal Q}}
\end{equation}
where $\rh$ denotes the DoS also called spectral function, ${\cal Q}$
denotes conserved quantities (quantum channel), $p=(p_0,\p)$ is the
total energy-momentum of the state. $V$ is the volume, and we will use
finite volume normalization for the states.

In order to be able to study the transport coefficients in general, we
define the correlators $C$ for generic conserved current $J$, and the
transport coefficient $D$ by the Kubo formula (linear response theory):
\begin{equation}
  C(x) = \exv{[J_i(x),J_i(0)]} \qquad D = \displaystyle
    \lim\limits_{\omega\to0} \frac {C(\k=0,\omega)}{\omega},
\end{equation}
for the shear viscosity $D=\eta$ and $J_i\to T_{12}$. The expectation
value is taken in equilibrium:
\begin{equation}
  \label{exvdef}
  \exv{A} =\frac1{\cal Z} \Tr e^{-\beta H} A,\qquad {\cal Z} = \Tr
  e^{-\beta H} = e^{-\beta F},
\end{equation}
where $F$ is the free energy.

To find out a generic formula for the transport coefficients we insert
an energy-momentum eigenbasis into the formula, and use the fact that
translation is generated by $P_\mu$ energy-momentum operator as
$J_i(x) = e^{iPx} J_i(0) e^{-iPx}$. Then we find
\begin{eqnarray}
  && C(x) = \frac1{\cal Z} \sum\limits_n\exv{n|e^{-\beta H}
    [J_i(x),J_i(0)]|n} =\nn&& 
  = \frac1{\cal Z} \sum\limits_{n,m}\biggl(
    \exv{n|e^{-\beta H} J_i(x)|m}\exv{m|J_i(0)]|n} -\nn&&\hspace*{4em}
    \exv{m|e^{-\beta H} J_i(0)|n}\exv{n|J_i(x)]|m} \biggr) = \nn&&=
  \frac1{\cal Z} \sum\limits_{n,m} e^{i(P_n-P_m)x}  \left( e^{-\beta
      E_n} - e^{-\beta E_m} \right) |\exv{n|J_i(0)|m}|^2.
\end{eqnarray}
By inserting the DoS we find
\begin{eqnarray}
  C(x)&&=\frac{V^2}{\cal Z} \sum\limits_{\cal K,Q} \pint4k \frac{d^4
    q}{(2\pi)^4} \rh_{\cal K}(k) \rh_{\cal Q}(q)
  e^{i(k-q)x}\times\nn&&\quad \times  \left(
    e^{-\beta k_0} - e^{-\beta q_0} \right) |\exv{k,{\cal
      K}|J_i(0)|q,{\cal Q}}|^2,
\end{eqnarray}

After Fourier transformation, with $\p=0$ zero spatial momentum
\begin{eqnarray}
  &&C(\omega,\p=0) = \frac{V^2}{\cal Z} \sum\limits_{\cal K,Q} \pint4k
  \rh_{\cal K}(k) \rh_{\cal Q}(\k,k_0+\omega)\times\nn&& \times \left(
    e^{-\beta k_0} - e^{-\beta (k_0+\omega)} \right) |\exv{\k,k_0,{\cal
       K}|J_i(0)|\k,k_0+\omega,{\cal Q}}|^2.
\end{eqnarray}
We can take into account that the current cannot change the quantum
numbers without changing the energy or momentum, then the diffusion
constant reads
\begin{equation}
  D = \beta \frac{V^2}{\cal Z} \sum\limits_{\cal K} \pint4k
  \rh^2_{\cal K}(k) e^{-\beta k_0}|\exv{k,{\cal K}|J_i(0)|k,{\cal K}}|^2.
\end{equation}
In the free theory the expectation value of the current can be written
in volume normalization as $\exv{k|J_i|q}=qk_i/(k_0 V)$, where $q$ is
the charge carried by the current; for the shear viscosity $q\to
k_j$. In general the current matrix element is proportional to the
velocity $v_i=k_i/k_0$, and in volume normalization it is
inversely proportional to the volume. So the generic formula is
proportional to the free case, and we can write
\begin{equation}
  |\exv{k,{\cal K}|J_i(0)|k,{\cal K}}|^2 = {\cal J}^2_{\cal K}(k^2)
  \left(\frac{qk_i}{k_0 V}\right)^2,
\end{equation}
where ${\cal J}_{\cal K}(k^2)$ is a nonperturbative correction
factor. Then
\begin{equation}
  D = \frac\beta{\cal Z} \sum\limits_{\cal K} \pint4k \frac{q^2
    k_i^2}{k_0^2} e^{-\beta k_0} \left(\rh_{\cal K}(k) {\cal J}_{\cal
      K}(k^2) \right)^2.
\end{equation}
As a final step we can average over the spatial angular
dependence. This gives $\overline{k_i^2}= \k^2/3$ and $\overline{k_x^2
  k_y^2} = (\k^2)^2/15$. Then the shear viscosity reads
\begin{equation}
  \eta =  \frac\beta{15\cal Z} \sum\limits_{\cal K} \pint4k
  \frac{(\k^2)^2}{k_0^2} e^{-\beta k_0} \left(\rh_{\cal K}(k) {\cal
      J}_{\cal K}(k^2) \right)^2.
\end{equation}

After having defined the shear viscosity, we can try to give a formula
for the entropy density. This is, however, a conceptually more
difficult task. The point is that entropy density, being a statistical
concept, is not sensible in the microscopic theory. Indeed, if we
consider the volume dependence of the free energy defined from the
partition function in eq. \eqref{exvdef}, at small volumes it can show
any volume dependence (usually grows very fast), only at large volumes
will it be proportional to the volume. There is a ``crossover'' size
$L$, where statistical treatment starts to be sensible. This size
corresponds to the coarse graining scale, beyond that two neighboring
volume elements (grains) of linear size $L$ interact dominantly
through the surface, which modify the total energy of them only
weakly. This means that interactions are effectively cut off at scale
$L$, while statistics start to be valid above $L$. The value of $L$
can be read off the linear size of the cross section: for a strongly
interacting theory $L$ is large, a weakly interacting theory has small
$L$.

Having said that we choose a volume $V=L^3$ to define the free energy
density and entropy density as
\begin{equation}
  f = -\frac T{L^3} \ln\left(1+L^3\sum_{\cal K} \pint4k \rh_{\cal
      K}(k)\,e^{-\beta k_0}\right), \quad s=-\displaystyle\frac{\d
    f}{\d T}.
\end{equation}

So finally we obtain
\begin{equation}
  \frac\eta s= \frac{\displaystyle \frac\beta{15\cal Z}\sum_{\cal K}
    \pint4k \, \frac{(\k^2)^2}{k_0^2}\, e^{-\beta k_0}\, ({\cal J}_{\cal
      K}(k)\rh_{\cal K}(k))^2} {\displaystyle \frac\d{\d T} \frac
    T{L^3} \ln\left(1+L^3 \sum\limits_{\cal K} \pint4k \rh_{\cal
        K}(k)\,e^{-\beta k_0}\right)}.
\end{equation}

After having found this formula we can ask whether there is a lower
bound in this formula. The generic analysis can be found in
\cite{jak}, but the generic form can be easily understood. For that
consider the small entropy case: then in the log in the denominator
the $1$ will be dominant, and we can write
\begin{equation}
  \frac\eta s\biggr|_{\mathrm{small}\,s}= \frac{\displaystyle
    \frac\beta{15\cal Z}\sum_{\cal K} \pint4k \, \frac{(\k^2)^2}{k_0^2}\,
    e^{-\beta k_0}\, ({\cal J}_{\cal K}(k)\rh_{\cal K}(k))^2}
  {\displaystyle  \sum\limits_{\cal K} \pint4k
    \rh_{\cal K}(k)\,(1+\beta k_0) e^{-\beta k_0}}. 
\end{equation}
From here we see that the numerator is proportional to $\rh^2$ while
the denominator by $\rh$. If $\rh$ is zero almost everywhere and very
large at some point -- this is the small width quasiparticle limit --,
$\rh^2$ is even larger at these point, and the result will be
large. Therefore in small width quasiparticle limit $\eta/s$ is
large. On the other hand, is $\rh$ is small everywhere, then $\rh^2$
is even smaller, and the result is small. This means that small
$\eta/s$ can be expected in system where there are no long-lived
quasiparticles!

To be more quantitiative we rewrite our formula as:
\begin{equation}
  \label{exvratio}
  \frac{\eta}{s}\biggr|_{\mathrm{small}\,s} =
  \frac{\left\langle\left\langle \bar \rh^2 \right\rangle\right\rangle}
  {\left\langle\left\langle \bar \rh \right\rangle\right\rangle},
\end{equation}
where the rescaled DoS reads as
\begin{equation}
  \bar\rh_{\cal K}(k) = \frac{\beta}{15{\cal Z}} \frac{(\k^2)^2}
  {k_0^2 (1+\beta k_0)} {\cal J}^2_{\cal K}(k) \rh_{\cal K}(k),
\end{equation}
and the averaging is interpreted as $\left\langle\left\langle \dots
  \right\rangle\right\rangle = {\cal N}^{-1} \int dA(\dots)$ where the
integration measure is
\begin{equation}
  dA = 15{\cal Z} T \sum_{\cal K} \frac{d^4k}{(2\pi)^4} \frac{k_0^2(1+\beta
    k_0)^2} {(\k^2)^2} {\cal J}^{-2}_{\cal K}(k) e^{-\beta k_0},
\end{equation}
and the normalization factor reads
\begin{equation}
  {\cal N} =  15{\cal Z} T \sum_{\cal K} \pint4k
  \frac{k_0^2(1+\beta k_0)^2} {(\k^2)^2} {\cal J}^{-2}_{\cal K}(k)
  e^{-\beta k_0}.
\end{equation}
Now we can apply the Schwarz inequality: $\left\langle\left\langle
    \bar \rh^2 \right\rangle\right\rangle \ge \left\langle\left\langle
    \bar \rh \right\rangle\right\rangle^2$, and we see:
\begin{equation}
  \frac{\eta}{s}\biggr|_{\mathrm{small}\,s} \ge {\cal N}^{-1} s.
\end{equation}
What we see is that, although $\eta/s$ has a minimum, but it is no
universal (like $1/4\pi$), it is model- and environment-dependent. The
minimal value can even be zero by tuning the system to reach ${\cal
  N}\to\infty$ (cf. \cite{Cohan}), or go to $s\to0$ ie. to zero
temperature.

A more detailed analysis, including also systems with large entropy
density shows \cite{jak} that
\begin{equation}
  \frac{\eta}{s}\biggr|_{\mathrm{minimum}} = \frac{{\cal F}(L^3s)}
  {N_Q (LT)^4},
\end{equation}
where ${\cal F}(x)\sim x$ for small $x$ and $\sim e^x/x$ for large
$x$, and $N_Q$ is the effective number of quantum channels (particle
species).

\section{Model calculations}
\label{sec:2}

After having found the minimal value, we can try to construct
physically relevant models where this minimum can be reached. To this
end we make some simplifications in the above calculation. First, we
use a generalized quasiparticle systems, and substitute the DoS by the
quasiparticle spectral function. Then the free energy reads for
bosonic or fermionic systems
\begin{equation}
  \label{QPf}
  f = T \pint4k \,\rh_{QP}(k) \,(\mp)\ln\left( 1\pm e^{-\beta k_0}\right).
\end{equation}
A second simplifiacation is that we omit the radiative corrections
from the current matrix elements, ie. ${\cal J}_{\cal K}(k)=1$. Third,
we take only one quantum channel. In this way the ``reduced''
viscosity coefficient reads
\begin{equation}
  \label{QPeta}
  \bar\eta = \frac{\beta}{15}\,\pint4k \, \frac{(\k^2)^2}{k_0^2}\,
  e^{-\beta k_0}\, \rh_{QP}^2(k).
\end{equation}

\subsection{Small width case}

To check the validity of our approximations, we apply it to a
small-width quasiparticle case. Assume that the lowest lying states
can be approximated with Breit-Wigner form
\begin{equation}
  \rh_{QP}(k) = \frac{2\Gamma}{(k_0-\ep_k)^2 + \Gamma^2},
\end{equation}
where $\ep_k$ is the dispersion relation.

In the small width limit $\rh(q)^2 \approx \frac2\Gamma\,
2\pi\delta(q_0-\ep_q)$. The formulae \eqref{QPf} and \eqref{QPeta} can
be calculated \cite{jak} with the result
\begin{equation}
  \frac{\bar\eta}{s} = \frac T\Gamma f(\frac m T),
\end{equation}
where the function $f$ depends on the dispersion relation, and $m$ is
the mass scale. For example for $\ep_k=k,\; f = 540/\pi^4$ for bosons,
$f=4320/(7\pi^4)$ for fermions; if $\ep_k = m+\frac{k^2}{2m},\; f =
30\pi T/m$.

If we consider a massless theory (eg. conformal field theory) then the
only scale is $T$, and so $\Gamma\sim T$. This means that
$\bar\eta/s=$constant. The lower limit of this constant may come from
infinte coupling, $1/4\pi$.

In the massive case the width of the particle should come from
scattering process, therefore at low temperature we expect $\Gamma\sim
e^{-M/T}$, where $M$ is the energy of the lowest scattering state. In
this case $\bar \eta/s \sim T e^{M/T} \stackrel{T\to0}\to \infty.$

This means that in the small width quasiparticle case there is a lower
bound, which comes from the massless theory.

\subsection{Broad spectral function}

For an opposite case consider a flat spectral function:
\begin{equation}
  \rh_{QP}(k_0,k) = \frac{2\pi}{E_2-E_1}\Theta(E_1<k_0<E_2)
\end{equation}
which is a step function, where $E_{1,2}(k)=\sqrt{k^2+m_{1,2}^2}$. At
small temperatures ($T<m_1$) we find \cite{jak} 
\begin{equation}
  \frac{\eta}s = 6\pi \frac T{m_2-m_1}
\end{equation}
This formula suggests that by broadening the energy distribution, the
viscosity to entropy density ratio has no lower bound, it can be made
vanish. This is exactly the message of the generic analysis: since
$\rh$ is normalized, broadening means small values, and so $\rh^2\ll
\rh$ everywhere. This results in the smallness of $\eta/s$ ratio.

\subsection{System with zero mass excitations}

Although the above example may be not physical, there are examples
where the spectral function has no isolated quasiparticle peak, but it
starts with a continuum. This is the case when an interacting system
contains zero mass particles: then the multiparticle cut start
directly from the quasiparticle peak. The interpretation is that a
charged object is always surrounded by soft gauge bosons. For example
in QED at one loop level the electron spectral function
\cite{BogShir}, up to log corrections, is proportional to $\rh_{QP}(k) \sim
(k^2-m^2)^{-1}$, ie. it is divergent at the mass shell, and fall as
$1/(\omega-\ep_k)$. By Bloch-Nordsieck resummation it changes to
$1/(\omega-\ep_k)^{1-b}$, where $b$ is the beta function.

In order $\rh$ to be normalizable, the singular behaviour at the mass
shell must be smoothed out by some physical process (cf. also
\cite{Biro}). Near the threshold (which is the dominant regime at low
temperatures) we approximate
\begin{equation}
   \rh_{QP}(q) = {\cal C} q_0 \Theta(q-M) (q^2-M^2)^w,
\end{equation}
where $w\ge-1$ for normalizability, and $C$ is a constant, its
dimension is $[{\cal C}]=[E]^{-2(1+w)}$. Since $\eta\sim {\cal C}^2$
and $f\sim{\cal C}$, therefore ${\cal C}$ remains in the ratio. In the
massive and massless case we find
\begin{equation}
   \frac{\bar \eta}s \sim {\cal C}M^wT^{2+w}\quad\mathrm{and}\quad
   {\cal C} T^{2(1+w)} \stackrel{T\to0}{\longrightarrow}\quad 0.
\end{equation}
In the limiting $w=-1$ case the $\eta/s$ is constant, this is the case
in some conformal models.

This analysis suggests that in a system with zero mass exciations, but
normalizable spectral functions the $\eta/s$ ratio should be vanishing
at very small temperatures. 

\section{Conclusions}
\label{sec:concl}

In this paper we discussed the behaviour of the the shear viscosity to
entropy ratio in systems where the small-width quasiparticle
approximation is not necessarily true. We have found that, although
$\eta/s$ has a lower bound in each given system at a given
temperature, but this bound is system- and temperature dependent; for
small entropy case we found for the minimum $\sim \frac{s}{N_Q LT^4}$
where $L$ is the interaction length and $N_Q$ is the number of
effective quantum channels (particle species). This minimum can be
smaller than $1/4\pi$. This fact was demonstrated by constructing
models with this property: if the spectral function does not go to an
isolated quasiparticle peak at zero temperature, which is a natural
setup in systems with zero mass excitations, then $\eta/s\to0$ at zero
temperature.

\section*{Acknowledgment}

This work was supported by the Hungarian Science Fund (OTKA) K68108.

\end{document}